\input psfig
\input harvmac
\newcount\figno
\figno=0
\def\fig#1#2#3{
\par\begingroup\parindent=0pt\leftskip=1cm\rightskip=1cm\parindent=0pt
\global\advance\figno by 1
\midinsert
\epsfxsize=#3
\centerline{\epsfbox{#2}}
\vskip 12pt
{\bf Fig. \the\figno:} #1\par
\endinsert\endgroup\par
}
\def\figlabel#1{\xdef#1{\the\figno}}
\def\encadremath#1{\vbox{\hrule\hbox{\vrule\kern8pt\vbox{\kern8pt
\hbox{$\displaystyle #1$}\kern8pt}
\kern8pt\vrule}\hrule}}

 \overfullrule=0pt

\Title{\vbox{
\hbox{HUTP-01/A010}
\hbox{RUNHETC-2001-07}
\hbox{\tt hep-th/0103011}}}{On Domain Walls of}
\vskip-1cm
\centerline{\titlefont ${\cal N}=1$ Supersymmetric
Yang-Mills
in Four Dimensions}
\bigskip\bigskip

\centerline{Bobby Acharya $^1$ and
 Cumrun Vafa$^2$}
\bigskip
\centerline{$^1$ Department of Physics}
\centerline{ Rutgers University}
\centerline{Piscataway, NJ 08855-0849}
\bigskip
\centerline{$^2$ Jefferson Physical Laboratory}
\centerline{Harvard University}
\centerline{Cambridge, MA 02138, USA}
\bigskip

\vskip .3in
We study the BPS domain walls of supersymmetric Yang-Mills
for arbitrary gauge group $G$. We describe
the degeneracies of domain walls interpolating
between arbitrary pairs of vacua.  A recently proposed large $N$ duality
sheds light on various aspects of such domain walls.  In particular,
for the case of $G=SU(N)$
 the domain walls correspond to wrapped
$D$-branes giving rise to a 2+1 dimensional
$U(k)$ gauge theory on the domain wall 
with a Chern-Simons term at level
$N$.
This leads to a counting of BPS degeneracies of domain
walls in agreement with expected results.
\Date{March 2001}
\newsec{Introduction}

${\cal N}=1$ Yang-Mills in $d=4$ for a gauge group $G$ admits
$V=c_2(G)$ vacua where $c_2(G)$ denotes the dual Coxeter
number of the group.  In particular the $U(1)$ R-symmetry
is anomalous, having a ${\bf Z_{2V}}$ anomaly free subgroup, which
is spontaneously broken to ${\bf Z_2}$ by the gaugino condensate
$$\langle \lambda \lambda  \rangle =\Lambda ^3 \omega$$
where $\omega ^{V}=1$.  This raises the possibility
of having domain walls.
In fact one can consider BPS saturated domain walls
since the value of the superpotential in each vacuum is proportional
to the gaugino condensate.  In particular the BPS tension of a domain
wall connecting a pair of vacua which are separated by $k$ units in  phase
of the gaugino condensate
is given by 
$$T=|\Delta W| \sim \Lambda^3 | (1-\omega^k)|$$
The basic aim of this paper is to study the existence
and degeneracy of domain walls which interpolate between any pair of
such vacua, a question which clearly depends on $k$ and the group $G$.

A related problem is the study of domain walls in
${\cal N}=1$ supersymmetric Wess-Zumino model in 4 dimensions.
The Wess-Zumino model with a superpotential $W$ has as many vacua
as the number of critical points of $W$, i.e. solutions to
$dW=0$ and we can consider domain walls interpolating
between these vacua.  This is equivalent to counting
the kink solutions in its dimensional reduction to 2 dimensions
(in which case it is often referred to as the ${\cal N}=2$ Landau-Ginzburg model).
In this context, the question of the number of BPS kinks (weighted with suitable
signs \ref\cfiv{S.~Cecotti, P.~Fendley, K.~Intriligator and C.~Vafa,
Nucl.\ Phys.\ B {\bf 386}, 405 (1992)
[hep-th/9204102]} )
has been completely
solved \ref\cecv{S.~Cecotti and C.~Vafa,
Commun.\ Math.\ Phys.\ {\bf 158}, 569 (1993)
[hep-th/9211097].}\
and is related to intersection
theory in the context of the Picard-Lefshetz theory of singularities.

\lref\cecova{S.~Cecotti and C.~Vafa,
Phys.\ Rev.\ Lett.\ {\bf 68}, 903 (1992)
[hep-th/9111016] }

In fact the two problems are not unrelated:  Consider
the case of gauge group $G=SU(N)$.  In this case,
as noted in 
\refs{\cecv ,\cecova }\
the effective superpotential
for the gaugino superfield, is identical to that of the supersymmetric
sigma model on ${\bf CP}^{N-1}$,  in terms of linear
sigma model fields \ref\witl{E. Witten, Nucl. \ Phys.\
B {\bf 403} 159 (1993) [hep-th/9301042].}.  In particular
both
have
$N$ vacua.  The number of BPS domain walls interpolating
between vacua separated by $k$ units for the ${\bf CP}^{N-1}$ sigma
model is known to be $N!/k!(N-k)!$ \ref\hiv{K.~Hori, A.~Iqbal and C.~Vafa,
hep-th/0005247.}
, and this suggests the same
answer for the $SU(N)$ case.  In fact this can be made even more plausible
by noting that the compactification of the $SU(N)$
theory on the $S^1$
gives
rise to a superpotential \ref\kav{S.~Katz and C.~Vafa,
Nucl.\ Phys.\ B {\bf 497}, 196 (1997)
[hep-th/9611090].}\ which is exactly
the same as the {\it mirror} description for the ${\bf CP}^{N-1}$ discovered
in \ref\hv{K.~Hori and C.~Vafa,
hep-th/0002222.}.  Or if we compactify the 4d theory on $T^2$ the
theory has a branch which corresponds
to a sigma model on the moduli space of flat $SU(N)$ connections
on $T^2$,
which is ${\bf CP}^{N-1}$ \lref\looj{E. Looijenga, Invent. Math. {\bf 38} (1977) 17;
Invent. Math. {\bf 61} (1980) 1.}\lref\fmw{R. Freedman,
J. Morgan and E. Witten, Comm. Math. Phys. {\bf 187 } 679
(1997)[hep-th/9701162].} \refs{\looj , \fmw}.

In recent years there has been a fairly subtantial
amount of work focussing on BPS domain walls in the
${\cal N}$ $=1$ Yang-Mills theory from the viewpoint of low energy effective
field theory
\ref\d1{G.~Dvali and M.~Shifman,
Phys.\ Lett.\ B {\bf 396}, 64 (1997)
[hep-th/9612128]\semi A.~Kovner, M.~Shifman and A.~Smilga
Phys.\ Rev.\ D {\bf 56}, 7978 (1997)
[hep-th/9706089]\semi A.~Smilga and A.~Veselov,
Phys.\ Rev.\ Lett.\ {\bf 79}, 4529 (1997)
[hep-th/9706217]\semi A.~V.~Smilga and A.~I.~Veselov,
Phys.\ Lett.\ B {\bf 428}, 303 (1998)
[hep-th/9801142]\semi A.~V.~Smilga,
Phys.\ Rev.\ D {\bf 58}, 065005 (1998)
[hep-th/9711032]\semi M.~Shifman,
Phys.\ Rev.\ D {\bf 57}, 1258 (1998)
[hep-th/9708060]\semi M.~A.~Shifman and M.~B.~Voloshin,
Phys.\ Rev.\ D {\bf 57}, 2590 (1998)
[hep-th/9709137]\semi I.~I.~Kogan, A.~Kovner and M.~Shifman,
Phys.\ Rev.\ D {\bf 57}, 5195 (1998)
[hep-th/9712046]\semi T.~Matsuda,
[hep-th/9805134]\semi T.~Matsuda,
Phys.\ Lett.\ B {\bf 436}, 264 (1998)
[hep-ph/9804409]\semi A.~Campos, K.~Holland and U.~J.~Wiese,
Phys.\ Rev.\ Lett.\ {\bf 81}, 2420 (1998)
[hep-th/9805086]\semi M.~Shifman,
[hep-th/9807166]\semi A.~V.~Smilga,
[hep-th/9807203]\semi G.~Dvali, G.~Gabadadze and Z.~Kakushadze,
Nucl.\ Phys.\ B {\bf 562}, 158 (1999)
[hep-th/9901032]\semi B.~de Carlos and J.~M.~Moreno,
Phys.\ Rev.\ Lett.\ {\bf 83}, 2120 (1999)
[hep-th/9905165]\semi B.~de Carlos and J.~M.~Moreno,
[hep-th/9910208]\semi V.~S.~Kaplunovsky, J.~Sonnenschein and S.~Yankielowicz,
Nucl.\ Phys.\ B {\bf 552}, 209 (1999)
[hep-th/9811195]\semi Y.~Artstein, V.~S.~Kaplunovsky and J.~Sonnenschein,
JHEP{\bf 0102}, 040 (2001)
[hep-th/0010241].}.
A summary of many of the properties of the solutions
can be found in \ref\d19{D.~Binosi and T.~ter Veldhuis,
[hep-th/0011113]\semi B.~d.~Carlos, M.~B.~Hindmarsh, N.~McNair and J.~M.~Moreno,
hep-th/0102033.}.  
One of the important features of the BPS domain walls is that they behave like
$D$-branes for the QCD string as can be seen
from the $M$ theory formulation of ${\cal N}=1$ Yang-Mills\lref\witm{E.~Witten,
Nucl.\ Phys.\ B {\bf 507}, 658 (1997)
[hep-th/9706109].}\lref\sra{S. Rey, unpublished.} 
\refs{\witm, \sra}.

As for the degeneracy of the $SU(N)$ domain walls between vacua
separated by one unit of the gaugino condensate phase, the embedding
into $M$ theory \witm\ resulted in finding a single domain wall
\ref\volv{A.~Volovich,
Phys.\ Rev.\ D {\bf 59}, 065005 (1999)
[hep-th/9801166].} .  This appears to be in contradiction with the expected
answer of $N$, noted above,  at least for the domain walls connecting
adjacent vacua.  We will be able to shed light on this apparent
contradiction in this paper.

More recently, certain large $N$ dualities for ${\cal N}=1$
Yang-Mills were proposed \lref\vaug{C.~Vafa,
hep-th/0008142}\lref\ks{I.~R.~Klebanov and M.~J.~Strassler,
JHEP{\bf 0008}, 052 (2000)
[hep-th/0007191]}\lref\mnu{J.~M.~Maldacena and C.~Nunez,
Phys.\ Rev.\ Lett.\ {\bf 86}, 588 (2001)
[hep-th/0008001].} \refs{\vaug ,\ks ,\mnu }.  In particular the duality
proposed in \vaug\ was reformulated in terms of $M$ theory on $G_2$ holonomy
manifolds in \lref\ach{B.~S.~Acharya,
hep-th/0011089.}\lref\amv{M.~Atiyah, J.~Maldacena and C.~Vafa,
hep-th/0011256.} \refs{\ach ,\amv}\ and
explained in terms of a geometric flop in \amv .

We will use these recently discovered dualities to study
the ${\cal N}=1$ Yang-Mills domain walls.
Our main insight
will be to realise that the world-volume theory on the
domain walls in these large $N$ duals is
a particular supersymmetric Chern-Simons-Yang-Mills
theory. The enumeration of BPS domain walls is
thus reduced to appropriately
counting supersymmetric vacua of this theory.
We will be able to recover the results anticipated based upon
the equivalance with the supersymmetric ${\bf CP}^{N-1}$ sigma model.
Moreover we explain the apparent discrepancy with the result based on
$M$QCD, by noting that the issue involves global boundary conditions on
the wall - counting vacua in a toroidally compactified theory is not
necessarily the same as counting vacua in Minkowski space.

The organization of this paper is as follows:  In section
2 we discuss the prediction of the number of domain
walls interpolating between adjacent vacua using
the dual $G_2$ holonomy geometry, as well as from the
Type IIA description involving RR-flux.   In section 3 we use the
Type IIA superstring description of the domain wall
to compute the degeneracies of all the domain walls
for arbitrary separation of vacua. In particular we show that
the theory on the domain wall interpolating between vacua separated by
$k$ units, is an ${\cal N}=1$
 $U(k)$ gauge theory, with level $N$ Chern-Simons
term, with an extra scalar adjoint field.
In section 4 we find the degeneracies of domain
walls for arbitrary group $G$ by considering the compactification
of the theory to 3 and 2 dimensions and using the dual
Landau-Ginzburg description.
This turns out to provide a very simple general answer for
the degeneracies of BPS domain walls for
all gauge groups, based on the Dynkin diagram and the Dynkin numbers
of the gauge group.
  We also comment on how these general results might be
derived in the context of $G_2$ holonomy geometries dual
to ${\cal N}=1$ systems.

\newsec{Domain Walls of ${\cal N}=1$ Yang-Mills}

In the $M$ theory formulation \refs{\ach,\amv}\ of the duality of \vaug
, one considers a $G_2$ holonomy geometry, which is topologically
$({S^3}\times {\bf R^4}){\bf /Z_N}$, where ${\bf Z_N}$ acts differently
in the IR versus the UV region, in a continuous way \amv .
We will be mainly interested in the IR description where ${\bf Z_N}$
acts on $S^3$ freely and produces a Lens space.  From the Type IIA
perspective, if we choose the eleventh dimension to be the fiber of the
Hopf map ${S^3}/{\bf Z_N}$ $\rightarrow$ $S^2$
this corresponds to an $S^2$ geometry
with $N$ units of RR-flux through it.  The domain wall corresponds
to an $M5$-brane wrapped over $S^3/{\bf Z_N}$, which in Type IIA theory is realized
by a $D4$-brane wrapped over $S^2$ (with $N$ units of RR flux through
it).  Each such domain wall shifts the vacuum by one unit, as
is evident from the identification of the supersymmetric
vacua with the geometry \vaug .

Let us consider the theory on a single $M$5-brane at low energies. This theory
is free and in particular contains
a 2-form field, $b$. If we wrap the $M$5-brane on a manifold
$M$, then there arises the possibility of turning on a topologically
non-trivial, but {\it flat} $b$-field on $M$. Such choices are classified by
$H^2(M,U(1))$ and these $b$-field backgrounds are the analogue of backgrounds with
discrete torsion
in string theory
\lref\cvd{C.~Vafa,
Nucl.\ Phys.\ B {\bf 273}, 592 (1986).}\lref\esharp{E.R. Sharpe,
hep-th/0008191 .} \refs{\cvd ,\esharp}.  However
$H^2(S^3/{\bf Z_N},
U(1))$ is trivial, so we do not have the freedom to turn on any discrete
torsion.  In other words there is a single domain wall in $3+1$ dimensions.
Let us now consider compactifying one spatial direction on a circle.
In this case we can turn on discrete torsion along the torsion class
2-cocycles formed from the 1-cocycle
on the circle and the non-trivial 1-cocycles in $H^1({S^3/{\bf Z_N}}, U(1))$ since,
$$H^2(S^3/{\bf Z_N} \times S^1) \cong {\bf Z_N}$$
Thus upon going to 2+1 dimensions on a circle, we do get $N$ BPS domain
walls interpolating between adjacent vacua, as expected from considering
the solitons of the ${\bf CP}^{N-1}$ supersymmetric sigma model.
If we consider the circle going down to 3 dimensions as the
``11''-th direction, then from the perspective of Type IIA theory
the domain wall corresponds to a $D4$-brane wrapped over $S^3/{\bf Z_N}$
and in that case we can turn on any of the $N$ inequivalent
flat connections on the brane world-volume \ref\gopval{R.~Gopakumar and C.~Vafa,
Adv.\ Theor.\ Math.\ Phys.\ {\bf 2}, 399 (1998)
[hep-th/9712048].}.  These correspond to the $N$ choices of discrete
torsion in the $M$ theory perspective.

\subsec{Type IIA perspective in 3+1 dimensions}

There is yet another way to understand this
domain wall degeneracy, and that is
from the perspective of Type IIA theory where instead we now consider
the ``11-th'' direction to be the fibers of the Hopf map
${S^3}/{\bf Z_N} \rightarrow S^2$.
The wrapped $M$5-brane domain wall goes over
to a $D4$-brane wrapped around $S^2$.
The existence of the $U(1)$ gauge field on the wrapped $D$-brane
is related to the fact that the fundamental string can
end on the $D$-brane and thus provides a source for it.
Since the domain wall is a wrapped $D$-brane
and the QCD string in this
context is identified with the fundamental string,
this is consistent with the observation \witm,\sra\
that the QCD string can end on the Yang-Mills domain wall.

On
the world-volume
of the domain wall we have a $2+1$-dimensional $U(1)$ gauge theory with
${\cal N}$ $=$ $1$ supersymmetry, together with a scalar multiplet, giving
the normal position of the domain wall.  This can be viewed
as the multiplet we would have gotten if we did not have
any flux through the $S^2$, in which case we would have gotten
an ${\cal N}$ $=$ $2$ multiplet in $2+1$ dimensions on the domain wall.
However the flux on the domain wall breaks this to ${\cal N}$$=1$
in the following way:  The gauge field picks up a Chern-Simons
term of level $N$.  This follows from the fact that on the
$D4$-brane world-volume there is the interaction
$$\int {\tilde A}\wedge F\wedge F =\int {\tilde F} \wedge A\wedge dA$$

where ${\tilde A}$ is the bulk RR-gauge potential of Type IIA string theory and
$F$ denotes the field strength of the $U(1)$ gauge field $A$
on the $D4$-brane.  Since $\int{\tilde F}$ over $S^2$ is $N$, this gives
a Chern-Simons term of level $N$ (after one takes into account
the correct normalizations in the action). The existence of this interaction
is consistent with the fact that without RR-flux, the Type IIA theory has
twice the amount of supersymmetry, so the $D4$-brane in the theory without
flux would have ${\cal N}$ $=2$ supersymmetry on its world-volume. Turning on
RR-flux breaks the bulk Type IIA supersymmetry by half and that induces
the above Chern-Simons interaction on the $D4$-brane, which breaks its
supersymmetry by half also.  Note that if we turn on a field
strength $F$ on the domain wall, say as a delta function,
it serves as a source for the gauge
field $A$, with $N$ units of charge, due to this
Chern-Simons term.  This  charge
can be neutralized by $N$ fundamental strings ending on
the domain wall.  This is to be identified with
the Baryon vertex on the domain wall, and is consistent with the fact
that a $D2$-brane wrapped over $S^2$ can be viewed
as the Baryon vertex  (cf.  \ref\wbaryon{
E.~Witten,
JHEP{\bf 9807}, 006 (1998)
[hep-th/9805112].}, \ref\groog {D.~J.~Gross and H.~Ooguri,
Phys.\ Rev.\ D {\bf 58}, 106002 (1998)
[hep-th/9805129]} )
of the four dimensional gauge theory, which in turn is equivalent to turning on a delta
function flux for $F$ on the $D4$-brane worldvolume.

The $U(1)$ gauge theory on the $2+1$ dimensional
world-volume of the domain wall is massive, due to this interaction.
%
%
%
%
%
%
%
This mass is identified with the  only mass scale in the problem
which is the physical scale of ${\cal N}=1$ Yang-Mills.  Note that the
existence of the CS term and the generation of the mass
for the gauge field implies that there are no 
long range propagating
massless modes (other than the normal deformation of the domain
wall) on the domain wall.

Even though the gauge theory on the domain wall is massive, the
zero modes survive and lead to a topological counting of the number
of vacua \ref\wittencs{E.~Witten,
Commun.\ Math.\ Phys.\ {\bf 121}, 351 (1989)}.  This has been studied mostly in compact spaces, in
particular when we compactify the theory on the wall from $2+1$ to $0+1$
on a $T^2$
in which case we get $N$ vacua, i.e. the number of conformal blocks of RCFT,
for the $U(1)$ theory at level $N$. Note, that this corresponds to the
statement that in the four dimensional super Yang-Mills theory compactified
on $T^2$, there are $N$ domain walls between neighbouring vacua.

Even if we consider the theory
on $S^1$ instead of the torus,
the above results suggest that we should still
obtain $N$ vacua for the $U(1)$ Chern-Simons theory at level $N$.
We also note that since this $U(1)$ theory is free, the uncompactified
theory has one bosonic zero energy state,
consistent with the $M$ theory description above.

We have thus considered the number of domain walls between
adjacent vacua in supersymmetric $SU(N)$ gauge theory from various
points of view and also resolved the puzzle of their
degeneracies in connection with the $M$ theory realization
\volv .  It is natural to ask what happens when we consider
domain walls separating vacua which are not adjacent.  It turns out that
the last viewpoint, namely that of Type IIA in 3+1 dimensions
with $k$ D4 branes wrapping the $S^2$ is the most fruitful for this
question and
we consider that next.  However, we will make comments about the
description involving $M$ theory, or Type IIA, on
$G_2$-holonomy geometries at the end of section 4, after we describe
the expected result for domain wall degeneracies for
arbitrary gauge groups.

\newsec{Gauge Theory on the Domain Wall}

Consider the Type IIA background considered in
\vaug . The spacetime contains an $S^2$ with $N$ units of RR-flux through
it. Consider $k$ D4 branes wrapping this $S^2$. These $k$ $D4$-branes are the domain
wall which interpolates between vacuum $l$ and $l+k$ (mod $N$)
of the ${\cal N}$ $=1$ theory.

The discussion
in the previous section easily extends to the case of
more than one brane and we obtain a $U(k)$ gauge theory on the
$2+1$ dimensional 
world-volume of the domain wall with the same matter content
as an ${\cal N}$$=2$ supersymmetric theory (which has 4 supercharges) but
broken to an ${\cal N}$$=1$ supersymmetric theory, by a Chern-Simons
term for the gauge field at level $N$.  We can view this as an
${\cal N}$$=1$ supersymmetric $U(k)$ gauge theory with Chern-Simons
coupling of level $N$ coupled to a massless scalar multiplet in the adjoint
representation.  We would like to compute the number of ground
states of this theory.  To be more preciese we compute the number
of ground states of this theory compactified on $T^2$, and identify
it with the number of BPS domain walls interpolating between
vacua which are $k$ units apart.  To be more precise we will 
compute the net number of vacua of the theory
weighted by $(-1)^F$,
except for the overall $U(1)$ factor which gives for every ground state
an opposite statistic ground state due to the fermionic zero mode in the
matter multiplet. This is because the Witten index of the full $U(k)$ theory
is zero, since the ``central $U(1)$-theory'' is free.
This $U(1)$ corresponds to the usual center of mass motion
of the domain wall which gives it the correct spacetime supersymmetry multiplet
structure. If the index of the full $U(k)$ theory were non-zero
the domain wall would not be in a supermultiplet.
This is also related to computing $Tr(-1)^F F
$ \cfiv\ for each kink sector.

In other words we regard the ${\cal N}$$=1$
$U(k)$ gauge theory coupled to the adjoint scalar,
as being a
$$U(k)={U(1)\times SU(k)\over {\bf Z_k}}$$
gauge theory, and we will be interested in computing
the Witten index of this system, modulo the trivial doubling
(leading to zero) coming from the fermionic zero mode in the $U(1)$
factor.

Thus we have to find the number of ground states of the $U(1)$ theory
at level $N$ times that of the ${\cal N}=1$   $SU(k)$ system at level $N$ 
coupled to an adjoint scalar.  The number of ground states coming
from the $U(1)$ system (ignoring the fermion zero mode in the matter
adjoint) is $N$.  For the $SU(k)$ system, we have moduli
given by the vev of the adjoint scalar $\langle \phi
\rangle $.
 If we compactify the theory on a $T^2$ we
have
to integrate over all allowed moduli. 
For any
non-vanishing 
vev of the adjoint scalar the $SU(k)$ gauge symmetry
is broken to a factor, which includes at least one $U(1)$
given by the direction of the vev in the adjoint of $SU(k)$.      
Thus the corresponding contribution to Witten index vanishes from 
such points due to the extra fermionic zero mode in the matter system
along that $U(1)$ direction.  Thus the only point in moduli space
which can contribute to the index is the origin.

Consider adding a mass term for the $SU(k)$ components of
$\phi$
(this can be done, for
example by making the gauge theory an ${\cal N}=2$ system).
This changes the behavior of the system for large vevs of $\phi$
without affecting the theory near the origin of $\phi$.  This
could have potentially changed the Witten index, because
we are changing the behavior of the system at infinity in the field
space.  However, as we already argued, away from the origin there
is no contribution to Witten index, and so with this deformation
we would not be modifying the Witten index computation.

Thus we need to compute the Witten index for an ${\cal N}=2$
$SU(k)$ gauge theory at level $N$.  This has been done in
\ref\japa{K.~Ohta,
JHEP{\bf 9910}, 006 (1999)
[hep-th/9908120].}\ref\ami{O.~Bergman, A.~Hanany, A.~Karch and B.~Kol,
JHEP{\bf 9910}, 036 (1999)
[hep-th/9908075].} following the work of \ref\witincs{E.~Witten,
hep-th/9903005.}, and it is given by the number of level $N-k$ representations
of affine $SU(k)$ theory.  The answer is
$$I_{SU(k)}=Tr_{SU(k)}(-1)^F= {(N-1)!\over (k-1)!(N-k)!}$$
To find the total number of vacua we include the contribution
of the $U(1)$ piece, which gives a factor of $N$; taking the ${\bf Z_k}$
quotient which takes us from $U(1)\times SU(k)$ 
to the $U(k)$ theory, gives an additional factor of $1/k$ and we obtain
$$I'_{U(k)}={N!\over k! (N-k)!}$$
(where the prime denotes deleting the zero mode
from the $U(1)$ fermionic factor).  This answer is
exactly as expected for the net number of BPS domain
walls of the ${\cal N}$$=1$ $SU(N)$ gauge theory interpolating between
vacua $k$ units apart.

Note that the above answer applies to the four dimensional theory compactified
on $T^2$. The number of walls in the four dimensional theory in non-compact
${\bf R^4}$ is given by the
index of the three dimensional Chern-Simons theory on ${\bf R^3}$ which may not be the
same as the index on ${T^2}{\times}{\bf R}$,
 as we have seen already for the case
of $k$ $=$ $1$.

\newsec{The BPS Domain Walls for $d=4, {\cal N}=1$ Supersymmetric
Gauge Theory for
 Arbitrary Gauge Group}

In this section we expand upon the known results in the literature
and compute the (net number of) BPS domain walls for $d=4$,
${\cal N}$$=1$ Supersymmetric Gauge theories for arbitrary 
gauge group compactified on a circle or $T^2$ to $3$ or $2$ dimensions.
Before we explain the derivation, we state the result which is
surprisingly simple:
For each group $G$ consider the affine Dynkin diagram.  Associate
to each node a fermion with a ``$U(1)$ charge''
given by the corresponding Dynkin index.  Note that the sum
of the Dynkin indices is $c_2(G)$, i.e., the total number of vacua
of the ${\cal N}$$=1$ theory.  Consider all the possible
monomials of the fermions with a total ``$U(1)$ charge''
of $k$.  Then the net number of domain walls between
vacua separated by $k$ units (for $1<k<c_2(G)$) is given by
the net number of fermions with total charge $k$, where the
net number is counted relative to the $(-1)^F$.
Note that for the $SU(N)$ case, the affine Dynkin diagram
has $N$ nodes each with Dynkin index $1$, and this gives
the expected answer $(-1)^k N!/k!(N-k)!$ for the number
of domain walls separating vacua which are $k$ units apart.

Upon compactifying ${\cal N}$$=1$ supersymmetric Yang-Mills
theory on $S^1$ or $T^2$ one obtains an effective Landau-Ginzburg theory
described as follows:  Consider the affine Dynkin diagram of
the group $G$, and for each node introduce a chiral field $Y_i$,
whose imaginary part is periodic with period $2\pi $.
Consider a theory with $4$ supercharges given by a holomorphic
superpotential
$$W=\sum_i e^{-Y_i}$$
with the constraint that $\sum_i a_i Y_i =0$, where
$a_i$ are the associated Dynkin indices.
It is a simple exercise to show that this Landau-Ginzburg
theory has $c_2(G)\
=\sum{a_i}$ vacua, and that the value of the superpotential is
given by $W=r \omega$ for some real parameter $r$, where $\omega$
is a $c_2(G)$-th root of unity.

For the compactification
on a circle, this superpotential was derived in \kav
\ref\vano{C.~Vafa,
Adv.\ Theor.\ Math.\ Phys.\ {\bf 2}, 497 (1998)
[hep-th/9801139].}\ by embedding the system into $F$-theory, and from a field
theory analysis in \ref\ft1{N.~Seiberg and E.~Witten,
hep-th/9607163.}\ref\ft2{N.~M.~Davies, T.~J.~Hollowood, V.~V.~Khoze and M.~P.~Mattis,
Nucl.\ Phys.\ B {\bf 559}, 123 (1999)
[hep-th/9905015]}.  In compactification to
2 dimensions this has been argued to arise by noting
that the compactification of the gauge theory on $T^2$
leads to an ${\cal N}$$=2$ supersymmetric sigma model on the
moduli space of flat $G$-connections on $T^2$ which in turn is given
by a weighted projective space $WP^{a_i}$ where the weights
are given by the Dynkin indices of the corresponding affine
Dynkin diagram \refs{\looj , \fmw}. This corresponds to a
$U(1)$ linear sigma model with ${\cal N}=(2,2)$ in $d=2$
with $rank(G)+1$ fields of charges given by the indices $a_i$,
as shown in figure 1 (which is borrowed from \fmw ).
 Then, as discussed in
\hv\  mirror symmetry maps this sigma model to a Landau-Ginzburg
theory with the superpotential given above.

\midinsert
\centerline{\psfig{figure=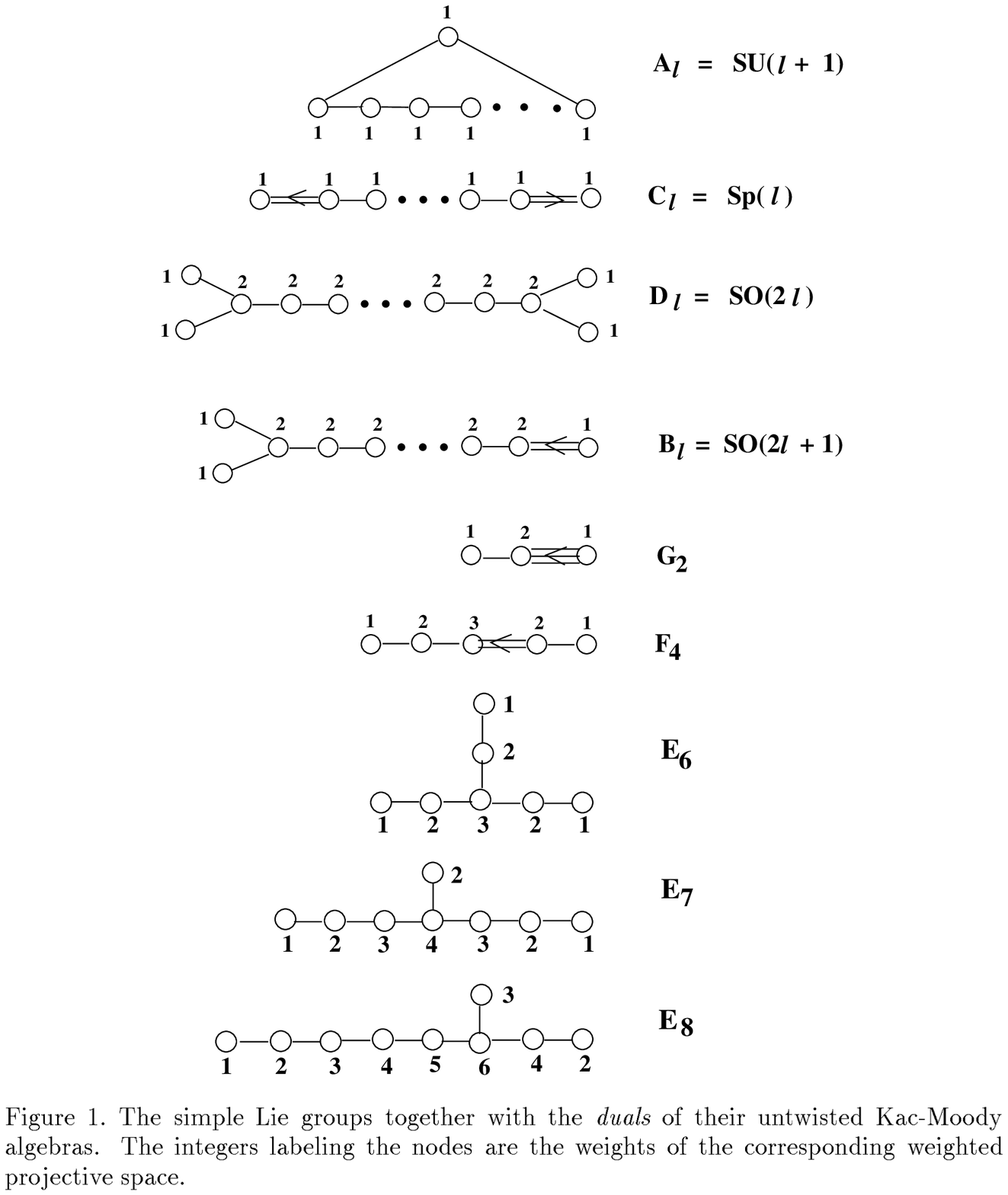,width=5.5in}}
\bigskip
\endinsert

The computation of the BPS domain wall solutions thus reduces
to the computation of the intersection number of vanishing cycles
of the corresponding Landau-Ginzburg theory, as discussed in detail
in \cecv .  In fact it is more convenient
to compute this intersection number by using the mirror symmetry
to the weighted projective space,
as was done in \hiv.  In this case one
considers a certain intersection number of $D$-branes
in the sigma model, which project to lines on the $W$ plane
emanating from the critical values and going to $W=+\infty$.
If we consider a sequence of such $D$-branes in an ascending
order (see figure 20 of \hiv ), the $l$-th $D$-brane corresponds
to the line bundle ${\cal O}(l)$ of the weighted projective
space $WP^{a_i}$.  Holomorphic sections of this bundle
correspond to symmetric monomials of total charge $l$ made
of the matter fields of the linear sigma model, which
have charges $a_i$.  Denote the number of such sections by $n_l$.
In fact this number also gives the index of the $\overline \partial$
operator coupled to the
${\cal O}(l)$ bundle, which in turn is the net number of ground states
of strings stretched between $D$-branes separated by $l$ units (as the other
cohomology groups of the $\overline \partial$-operator vanish).
Note that the generating function of this index is given by
$$S(q)={1\over \prod_i (1-q^{a_i})}=\sum_l n_l q^l.$$

Consider the upper triangular matrix $S=I+A
$ where $I$ is the identity matrix and $A$ is strictly
upper triangular with $A_{i,j}=n_{i-j}$, with $n_l$
as given by the above generating function.  Then the number
of kinks between the $i,j$ vacua is given by suitable brading
(discussed in \hiv ) which amounts to considering the inverse
matrix $S^{-1}$.  In particular $(S^{-1})_{ij}$ gives
the net number of kinks between the $i$-th and $j$-th vacua.  Clearly
the entries of $S^{-1}$ depend only on $i-j$.  If we encode
this information in a generating function $S^{-1}(q)$
then the condition that
this be inverse of $S$ implies that
$$S^{-1}(q)S(q)=1\rightarrow S^{-1}(q)=1/S(q)=\prod_i(1-q^{a_i})$$
This is exactly the partition function of $rank(G)+1$ fermionic
system graded by Dynkin indices.  Moreover the coefficient
of $q^k$ in the above expansion is the net number of kinks
between vacua separated by $k$ units, which leads us
to the statement that we made at the beginning of this section.

As an example consider the number of domain walls for the group
$E_7$. There are two BPS domain walls connecting
adjacent vacua corresponding to the two
nodes labeled by $1$ as can
be seen from figure 1 (each with fermionic number $(-1)^F=-1$).
For domain walls connecting the next nearest neighbor we have to consider
total Dynkin label of $2$.  This can be done either by taking
any of the three nodes labeled by $2$ or taking one copy of each
of the nodes labeled by $1$.  The first type will have fermion
number $-1$ and the second one will have fermion number $+1$
and so the net number of BPS domain walls interpolating
between next nearest neighbors
is given by $1-3=-2$.  Similarly
one can enumerate the net degeneracies for
all separations.  The total partition function of the domain
wall for $E_7$ gauge group is, as noted before, given by
$$(1-q)^2(1-q^2)^3(1-q^3)^2(1-q^4)$$
where the coefficient of $q^k$ is the net number
of domain walls separating vacua which are $k$ units apart
in the phase of the gaugino condensate.

\subsec{IIA on $G_2$-holonomy Perspective}

As noted in section 2, in the context of
the Type IIA theory we can also consider the 11-th
circle to correspond to the circle we choose to go from
4 to 3 dimensions.  In this way we have a Type IIA perspective
involving a background with a $G_2$-holonomy metric.  For
the $SU(N)$ group this contains the Lens space $S^3/{\bf Z_N}$.
The $D4$-branes wrapped over the Lens space correspond, as discussed
before, to domain walls interpolating between neighbouring vacua.  As noted
before there are exactly $N$ of them.  The question is whether
we can extend this to obtain also the result for the degeneracy
involving vacua separated by $k$ units.  In this case we 
consider $k$ $D$4 branes and get a $U(k)$
gauge theory on $S^3/{\bf Z_N}$.  Consider turning on flat connections
in $U(k)$.  If all the connections are inequivalent (which can
be done only if $k\leq N$) then the gauge group is broken down
to $U(1)^k$. If not the gauge group contains non-Abelian
factors. 
In general, there are scalar fields which are
massless classically. The question is whether there is a normalizable
ground state at the origin of field space. This can happen,
for example if the corresponding scalars pick
a mass (except for the overall $U(1)$).  This is in principle
allowed by the number of supersymmetries (namely (1,1) on the domain
wall in 1+1 dimension), but we have not demonstrated that it is generated.
If it is generated (or at any rate if there
is a normalizable zero mode at the origin), and in addition
if the configurations
with non-abelian factors, which arise
when some of the flat connections are taken
to be the same, make no contributions to the index,
this would explain
the result we obtained for the degeneracy of domain walls for
the $SU(N)$
case.  This sounds very much like the s-rule \ref\hanwit{A.~Hanany and E.~Witten,
Nucl.\ Phys.\ B {\bf 492}, 152 (1997)
[hep-th/9611230].}, and
is probably connected to it
by using some chain of dualities similar to the considerations of \japa .

We can also consider geometries involving $D$ and $E$ groups
(in principle we can also consider non-simply laced cases, 
for which the same arguments below should apply).  These correspond
to $G_2$ holonomy metrics having an $S^3/\Gamma$ where $\Gamma$ is
the associated group.  Aspects of the QCD strings for these
cases were discussed in \ref\achstr{B.~S.~Acharya,
hep-th/0101206.}.  Certain
aspects of the D case was discussed in \ref\siva{S.~Sinha and C.~Vafa,
hep-th/0012136.}.

As noted in \gopval\ for each irreducible
representation of $\Gamma$ of dimension $a_i$
we get a bound state of $a_i$ D4-branes wrapping over $S^3/\Gamma$.
Moreover the irreducible representations of 
$\Gamma$  are in 1-1 correspondence with the nodes of the affine
Dynkin diagram with $a_i$ corresponding to the Dynkin numbers.
Now if we consider $k$ D4 branes, we get a $U(k)$ system, and
decomposing it to flat connections using the irreducible
connections, this corresponds to splitting $k$ in terms of
sum of some number of $a_i$.  Again if all the flat
connections are inequivalent the gauge groups is broken to $U(1)^r$
for some $r$ and if our previous conjecture holds again, this
would explain the degeneracy we obtained above.

\vglue 1cm

We would like to thank K. Hori, A. Iqbal, D. Kabat, S. Katz,
H. Liu, G. Moore, A. Rajaraman, M. Rozali, M. Strassler and
D. Tong for valuable discussions.

\vglue 1cm

The research of CV is supported in part by NSF grants PHY-9802709
and DMS 9709694.

\listrefs

\end